\documentclass[]{spie}  

\usepackage{amsmath,amsfonts,amssymb}
\usepackage{graphicx}
\usepackage[colorlinks=true, allcolors=blue]{hyperref}

\title{Status and performance of the Asgard/NOTT cold nulling camera}

\author[a]{P. Chingaipe*}
\author[b]{G. Medgyesi}
\author[c]{A. Mazzoli}
\author[d]{E. Bouzerand}
\author[a]{R. Laugier}
\author[e]{P. Abraham}
\author[c]{C. Dandumont}
\author[a]{D. Defrère}
\author[d]{A. Glauser}
\author[d]{G. Garreau}
\author[f]{M. Ireland}
\author[e]{A.P. Joo}
\author[g]{J. Loicq}
\author[j]{M-A. Martinod}
\author[a]{K. Missiaen}
\author[h]{M. Montoya}
\author[a]{J. Morren}
\author[a]{G. Raskin}
\author[a]{M. Salman}
\author[f]{W. Schofield}
\author[a]{W. Verstraeten}
\author[h,l]{S. Ertel}
\author[h]{T. A. Stuber}
\author[h]{J. P. Scott}
\author[i]{S. Kraus}
\author[j]{F. Martinache}
\author[k]{X. Haubois}
\author[k]{N. Schuhler}

\affil[a]{Institute of Astronomy, KU Leuven, Celestijnenlaan 200D, 3001 Leuven, Belgium.}
\affil[b]{Fornax, H-1123 Budapest, Táltos u. 1., Hungary}
\affil[c]{Space sciences, Technologies \& Astrophysics Research (STAR) Institute, University of Li\`ege, Li\`ege, Belgium}
\affil[d]{ Eidgen\"ossische Technische Hochschule (ETH) Zurich, Institute for Particle Physics and Astrophysics, Zurich, Switzerland}
\affil[e]{Konkoly Observatory, HUN-REN Research Centre for Astronomy and Earth Sciences, MTA Centre of Excellence, Konkoly-Thege Mikl\'os \'ut 15-17, 1121 Budapest, Hungary}
\affil[f]{Research School of Astronomy and Astrophysics, Australian National University, Canberra, ACT 2611, Australia}
\affil[g]{Delft University of Technology, Netherlands}
\affil[h]{Department of Astronomy and Steward Observatory, 933 North Cherry Ave, Tucson, AZ 89 85721, USA}
\affil[i]{School of Physics and Astronomy, University of Exeter, Stocker Road, Exeter, EX4 4QL, United Kingdom}
\affil[j]{Université Côte d'Azur, Observatoire de la Côte d'Azur, CNRS, Laboratoire Lagrange, France}
\affil[k]{European Organisation for Astronomical Research in the Southern Hemisphere, Casilla, 19001, Santiago 19, Chile}
\affil[l]{Large Binocular Telescope Observatory, The University of Arizona, 933 North Cherry Avenue, Tucson, AZ 85721, USA}

\authorinfo{Further author information: *Peter Chingaipe, petermarley.chingaipe@kuleuven.be}

\begin{document} 
    \maketitle

    \begin{abstract}
   Long-baseline nulling interferometry suppresses the on-axis light of a star, enabling high-contrast observations of faint circumstellar emission and exoplanets at high angular resolution. The Nulling Observations of dusT and PlaneTs (NOTT) instrument is the \(L^{\prime}\)-band nulling interferometer of the Asgard instrument suite and will be the first nuller installed at the Very Large Telescope Interferometer (VLTI). Its main scientific goals are to characterise hot/warm exozodiacal dust and young Jupiter-like exoplanets at close (\(<10~\mathrm{AU}\)) orbital separations, where indirect detection methods are currently most sensitive. 
   \smallbreak
   NOTT uses a four-telescope integrated-optics photonic beam combiner operating over \(3.5{-}4.0~\mu\mathrm{m}\). The photonic chip produces photometric, bright, and nulled outputs that are relayed, polarization-split, spectrally dispersed, and imaged onto a \(5~\mu\mathrm{m}\)-cutoff HAWAII-2RG detector by the NOTT cold nulling camera. The cold optical assembly includes the photonic-chip interface, injection optics, collimator, Wollaston prism, filter-wheel-mounted dispersive element, imager, and cryogenic detector stage. The mechanical implementation is designed for operation near \(90~\mathrm{K}\), using aluminium support structures, spring-loaded optical mounts, and cryogenic-compatible mechanisms to preserve alignment during cooldown. 
   \smallbreak 
   Here, we present the current status and performance of the NOTT cold nulling camera and cryostat assembly. The system is undergoing assembly, integration, and verification to validate cooldown stability, thermal interfaces, detector operation, vacuum performance, and opto-mechanical repeatability. Initial cryogenic tests have demonstrated operation near \(100~\mathrm{K}\) for the cold optics and \(30~\mathrm{K}\) for the detector, cryogenic validation of the filter wheel, and a vacuum pressure of \(8 \times 10^{-8}~\mathrm{mbar}\). These activities represent a key step toward the deployment of NOTT as the first \(L^{\prime}\)-band nulling interferometer at the VLTI.
    \end{abstract}


    \section{INTRODUCTION}
    \label{sec:intro}  

        \begin{figure}[!ht]
    \centering
    \includegraphics[width=0.88\textwidth]{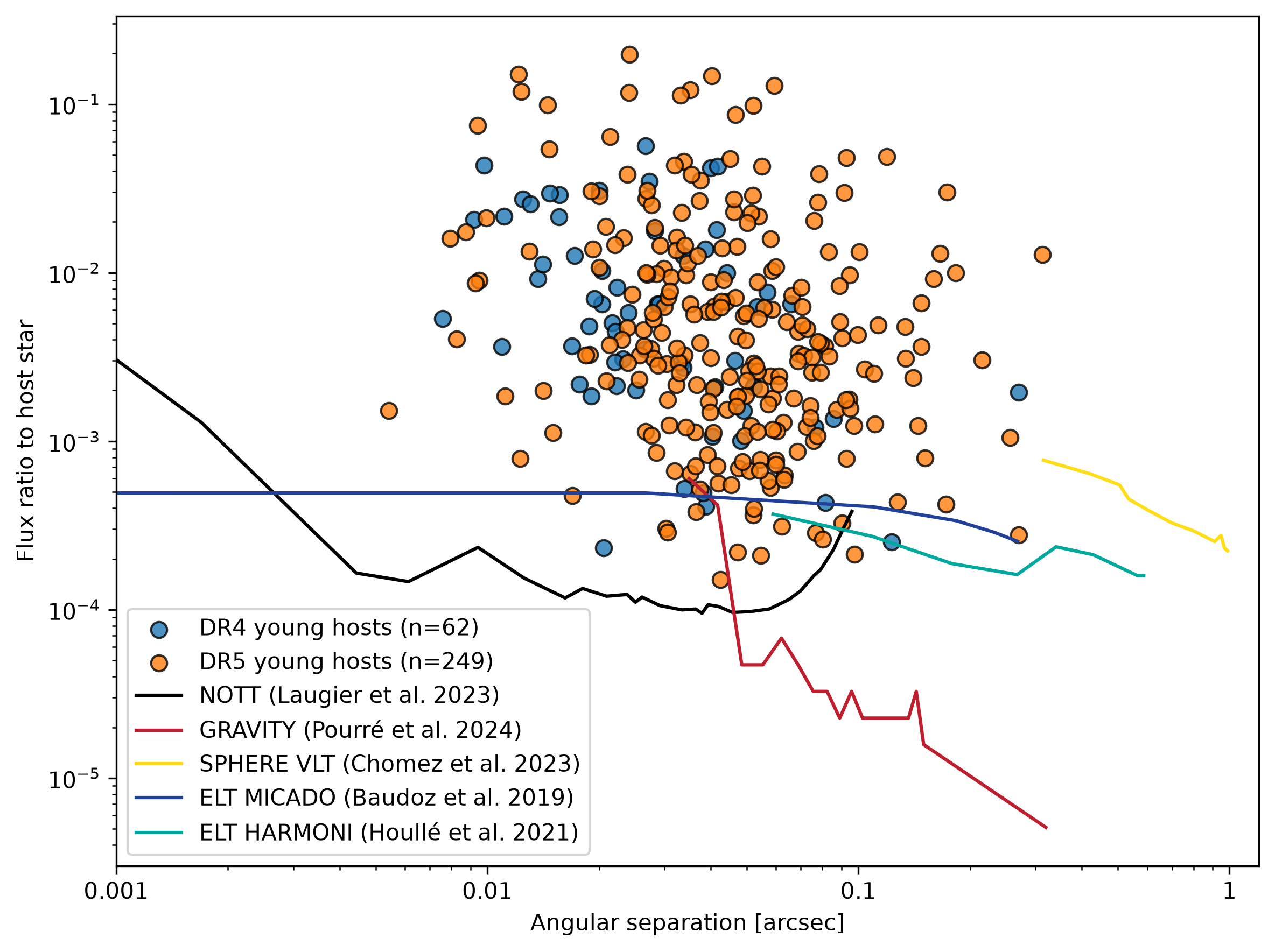}
    \caption{Predicted planet--star flux ratios as a function of angular separation for mock young-association planet hosts, compared with approximate GRAVITY and NOTT sensitivity curves. The mock targets are drawn from Lammers \& Winn (2025)\cite{2025LammersWinn} and restricted to \(\mathrm{Dec}<+16^\circ\) and distances within \(100~\mathrm{pc}\). The figure illustrates how \emph{Gaia} astrometry can identify candidate systems for NOTT follow-up in the relevant contrast--separation regime. Adapted from the DI-flux-ratio-plot script available at \url{https://github.com/nasavbailey/DI-flux-ratio-plot}.}
    \label{nott}
    \end{figure}

    To date, most directly imaged exoplanets have been discovered through “blind” surveys, where targets are selected based on age and proximity to Earth. These detections have primarily occurred in the near-infrared using 8-meter class ground-based telescopes equipped with extreme adaptive optics (AO) and coronagraphs. However, recent blind imaging campaigns such as Gemini-GPIES \cite{2019Nielsen} and SPHERE-SHINE \cite{2025Chomez} have yielded few new planet detections (see Fig.~\ref{nott}). For example, GPIES detected only nine companions from a sample of $\sim$300 stars, and three of those planets were already known prior to the survey.
     \smallbreak 
     Nulling interferometry is a powerful technique for directly imaging high-contrast exoplanets in a region of the parameter space that has so far only been covered by indirect detection methods. The high angular resolving power provided by mature infrastructure for long-baseline interferometry can aid this endeavour. These include the Very Large Telescope Interferometer (VLTI) and the Centre for High Angular Resolution Astronomy (CHARA), which routinely combine four and six telescopes, respectively. In a nuller, the on-axis stellar light is redirected toward one or more bright outputs, while light from an off-axis companion can be transmitted through the nulled outputs. 
     \smallbreak 
     With the long baselines (up to 130-m) of the four VLTI 8.2-m Unit Telescopes (UTs), previously-imaged planets can be well resolved. The first direct detection of an exoplanet by the GRAVITY instrument (HR 8799e; \cite{2019GRAVITY}) demonstrated that optical interferometry can achieve high-contrast imaging at milliarcsecond angular resolution. GRAVITY’s integrated optics beam combiner, operating in dual-field mode, enables both astrometry and $K$-band (2.0--2.4\,$\mu$m) spectroscopy of exoplanets. However, GRAVITY observations require prior knowledge of the planet’s location to position the second arm of the interferometer, and the dual-field design imposes an inner working angle (typically tens of milliarcseconds) that excludes the smallest separations --- where giant planets near the snow line reside \cite{2019Fernandes}. Despite these constraints, GRAVITY delivered groundbreaking results on many giant exoplanets, including precise astrometry and spectroscopic characterization.
     \smallbreak 
     In contrast to GRAVITY, Asgard/NOTT \cite{2018Defrere} ---the first nulling interferometer on the VLTI--- enables high-contrast \emph{on-axis} observations in the L$^\prime$-band (3.5--4.0\,$\mu$m), where the planet/star flux ratio is significantly improved compared to shorter wavelengths. Its primary scientific goals are to characterize young Jupiter-like exoplanets at close ($<10$ AU) orbital separations (see Fig.~\ref{nott}) and to detect hot/warm exozodiacal dust \cite{2022Defrere}. The timing of Asgard/NOTT aligns with a transformative phase in exoplanetary science: over the next few years, \emph{Gaia} is expected to deliver an abundance of giant planets in the 1--7 AU range \cite{2025LammersWinn}, enabling unprecedented studies of exoplanet demographics. Achieving this will require not only a deeper understanding of how to combine astrometry with direct imaging, but also advances in astronomical instrumentation. Here, we present the NOTT cold nulling camera, the cryogenic imager and spectrograph that records the spectrally dispersed outputs of the Asgard/NOTT photonic nuller \cite{sanny2026}.

\section{Instrument Context and Optical Path}
\label{sec:instr} 

\begin{figure}[!ht]
    \centering
    \includegraphics[width=0.88\textwidth]{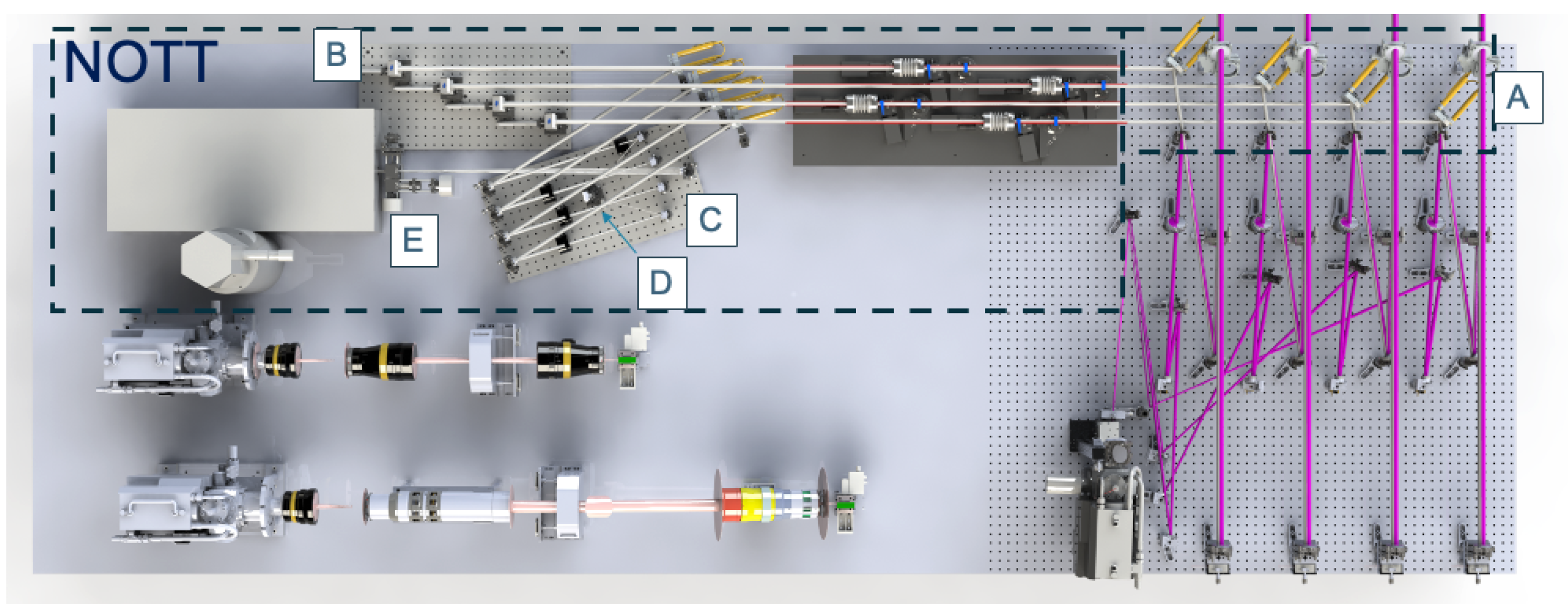}
    \caption{Top-view design of the Asgard instrument suite on the VLTI visitor table, showing the lower optical level. The NOTT warm-optics region is indicated by the dashed blue lines. After fringe tracking and wavefront correction by HEIMDALLR and Baldr, the four VLTI beams, shown in purple, are transmitted toward the NOTT beam path through CaF\(_2\) dichroics. Within the NOTT warm optics, each beam is steered by tip-tilt mirrors (A) for fine alignment and propagates along the edge of the optical table before reaching the delay lines (B), which provide optical-path-difference control. The beams are then directed toward the injection system (C) and pass through the pupil-recombination optics, where the slicer, a four-element mirror assembly (D), works with off-axis paraboloids to recombine the four pupils. The beams are finally injected toward the photonic chip and cryostat interface (E), which feeds the NOTT cold nulling camera. Adapted from Defr\`ere et al. (2024).}
    \label{asgard_layout}
    \end{figure}

    Asgard \cite{2023Martinod} is a four-module VLTI instrument suite composed of HEIMDALLR \cite{2018Ireland, 2026Martinache}, Baldr \cite{2024Courtney-Barrer}, BIFROST \cite{2024Kraus, 2026Kraus} and NOTT \cite{2018Defrere, 2022Defrere}. HEIMDALLR provides high-sensitivity fringe tracking, Baldr optimizes the wavefront with a Zernike wavefront sensor, BIFROST provides high-resolution spectro-interferometry, and NOTT performs high-contrast nulling interferometry. Figure.~\ref{asgard_layout} shows the opto-mechanical design of Asgard, in which the upstream modules stabilize the VLTI beams for science observations with BIFROST and NOTT.
    \smallbreak 
    The NOTT optical system receives the four VLTI beams from HEIMDALLR. As shown in Fig.~\ref{asgard_layout}, the NOTT warm optics correspond to the ambient-temperature optical elements in the NOTT beam path, upstream of the cryostat. Within these warm optics, each beam is controlled by two tip-tilt mirrors for fine beam alignment and one delay line for optical path difference control. The beams are then sent through the pupil-recombination optics, where the slicer---a four-element mirror assembly---and off-axis paraboloids recombine the four pupils before injection into the cryostat housing the NOTT cold nulling camera. 
    \smallbreak
    Because the NOTT cold nulling camera operates in the thermal infrared, its temperature requirements are set by the instrument background and detector performance. The cold optics are designed to operate near \(90~\mathrm{K}\), where their \(L^{\prime}\)-band emission remains below the allowed fraction of the telescope background. The HAWAII-2RG (H2RG) detector is required to operate below \(50~\mathrm{K}\), which increases the number of operable pixels, reduces persistence,\footnote{Persistence refers to residual signal remaining in detector pixels after previous illumination.} limits read-noise contributions from the ASICs, and improves the overall uniformity and usability of the detector array.

\section{The NOTT Cold Nulling Camera}
\label{sec:cam}
\begin{figure}[!ht]
    \centering
    \includegraphics[width=0.88\textwidth]{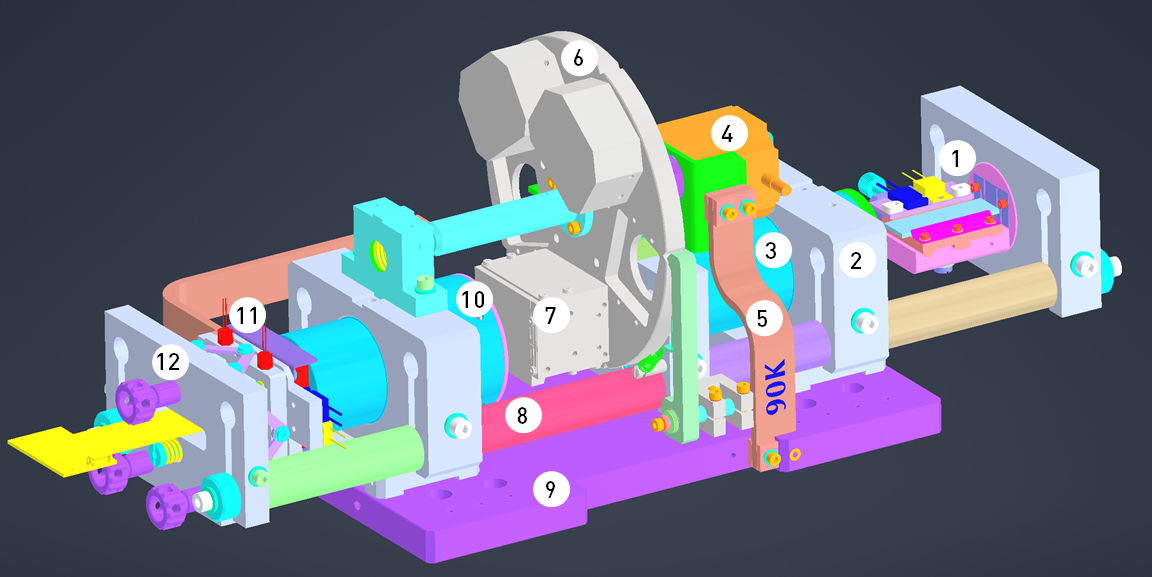}
    \caption{Mechanical design of the NOTT cold nulling camera by Fornax. The assembly includes the photonic-chip holder (1), vertical bar-clamping unit (2), collimator assembly (3), stepper motor (4), \(90~\mathrm{K}\) thermal strap (5), filter wheel (6), prism holder (7), horizontal bar (8), base plate (9), imager assembly (10), HAWAII-2RG detector (11), and detector moving stage (12).}
    \label{cold_camera_mechanics}
    \end{figure}

    \begin{figure}[!ht]
    \centering
    \includegraphics[width=1\textwidth]{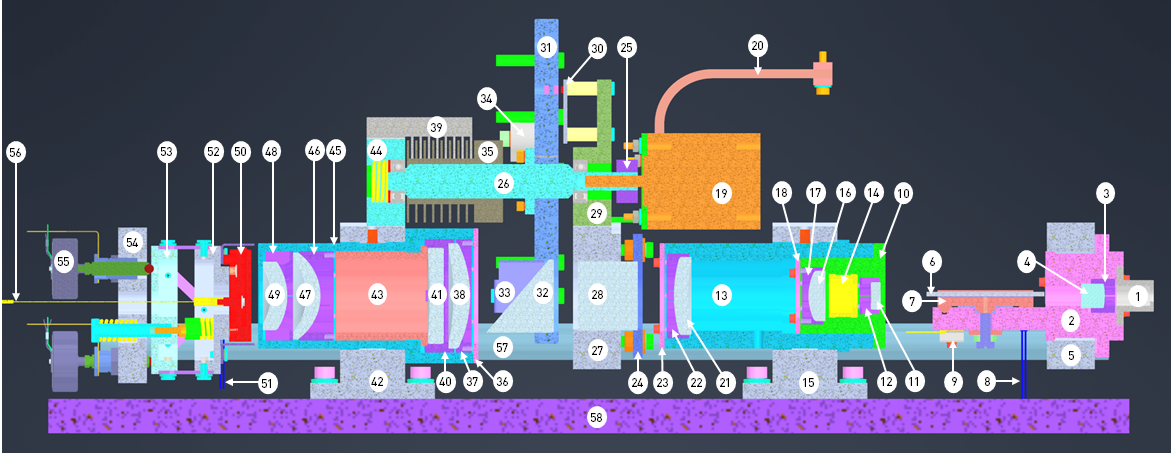}
    \caption{Annotated cross-section of the NOTT cold nulling camera mechanical assembly. The numbered components are: cover/cold stop (1), injection-lens tube (2), injection centering ring (3), injection lens (4), photonic-chip holder (5), NOTT GLS chip (6), chip holder (7), temperature sensor (8), heating-film fixing bridge (9), collimator tube 01 (10), collimator lens 01 (11), collimator centering ring 01 (12), collimator tube 02 (13), collimator cover 01 (14), vertical bar-clamping unit (15), collimator lens 02 (16), collimator centering ring 02 (17), collimator cover 02 (18), stepper motor (19), \(90~\mathrm{K}\) thermal strap (20), collimator lens 03 (21), collimator centering ring 03 (22), collimator cover 03 (23), Wollaston holder (24), clamp (25), filter-wheel shaft (26), Wollaston holder fix (27), Wollaston prism (28), bearing block (29), Hall sensor (30), filter wheel (31), prism (32), prism attachment (33), counterweight stack (34), cooling collar (35), imager cover 01 (36), imager centering ring 01 (37), imager lens 01 (38), cooling collar (39), imager centering ring 02 (40), imager lens 02 (41), vertical bar-clamping unit (42), spacer (43), bearing block (44), imaging tube (45), imager centering ring 03 (46), imager lens 03 (47), imager centering ring 04 (48), imager lens 04 (49), image plane (50), temperature sensor (51), insulated sensor holders (52, 53), sensor moving stage (54), sensor adjustment knob (55), sensor cable (56), horizontal shaft (57), and base plate (58).}
    \label{cold_camera_cross_section}
    \end{figure}

    Figure~\ref{cold_camera_mechanics} shows the mechanical design of the NOTT cold nulling camera by Fornax. The assembly implements the cold camera optical path downstream of the NOTT warm optics. All mechanical parts are made of aluminium, while the screws, springs, and washers are made of stainless steel. The supporting structure consists of two vertical bar-clamping units, which are rigidly bolted to the base plate, and two round horizontal bars. 
    \smallbreak 
    The aluminium support structure is dimensioned so that the optical geometry is correct at \(90~\mathrm{K}\). This was checked with a thermo-mechanical deformation analysis of the cold camera CAD model in Autodesk Inventor Nastran. Starting from the ambient-temperature geometry at \(293~\mathrm{K}\), the model was evaluated at \(90~\mathrm{K}\) to estimate the deformation after cooldown. The quantities checked were the relative positions of the aluminium mounting features (i.e., the centering and retaining rings shown in Fig.~\ref{cold_camera_cross_section}) that define the axial placement of the lenses. These were compared with the CODE V optical prescription determined at the University of Liège, which specifies the required temperature-dependent optical spacings. This comparison confirmed that the cooldown-deformed mounting geometry remains compatible with the required optical spacings at the cryogenic operating temperature. The check is important because the CODE V tolerancing shows sensitivity to axial lens positioning at the \(\sim10~\mu\mathrm{m}\) level if no compensation is made with detector focus. 
    \smallbreak
    As shown in the cross-section view in Fig.~\ref{cold_camera_cross_section}, the incoming beams enter the NOTT cold nulling camera through the cold stop, which defines the accepted pupil, and are focused by the \(12.0~\mathrm{mm}\)-diameter ZnSe injection lens onto the photonic chip. The chip is a four-input gallium lanthanum sulfide (GLS) chalcogenide-glass integrated-optics beam combiner based on a double-Bracewell architecture. It produces eight outputs: four photometric outputs, two bright outputs, and two nulled outputs. The characterized device demonstrated single-mode behaviour, achromatic directional couplers over \(3.65{-}3.85~\mu\mathrm{m}\), and a measured throughput of about \(37\%\) \cite{sanny2026}.
    \smallbreak
    The chip outputs then enter the collimator assembly, which houses three lenses: a \(12.0~\mathrm{mm}\)-diameter CaF\(_2\) lens followed by \(22.0~\mathrm{mm}\)- and \(34.0~\mathrm{mm}\)-diameter ZnSe lenses. Each lens is axially located by a flat reference surface in the lens mount and held in place by a flexible split centering ring. Small helical springs provide the preload against the chamfered lens edges.\footnote{The exact edge geometry follows the individual lens drawings; for the collimator lenses, the specified chamfered edge features are used for mechanical contact with the centering rings.} This mounting concept fixes the axial position of each lens through its flat seating surface, making the lens position insensitive to manufacturing tolerances of the lens or centering ring. The centering rings also maintain radial centering while allowing differential dimensional changes during cooldown, so the lenses are not subjected to excessive mechanical stress between ambient temperature and \(90~\mathrm{K}\).
    \smallbreak
    After collimation, the beams pass through a Wollaston prism mounted in an oversized holder to accommodate thermal-expansion differences. The prism is spring-loaded against its seating surface to maintain contact during cooldown. Since the prism has flat optical faces, its radial position is not critical. The main mechanical requirement is to maintain perpendicularity to the optical axis. In operation, the prism separates the two polarization states with a beam deviation of \(\pm0.02^\circ\). Splitting each of the eight photonic-chip outputs therefore results in 16 polarization-separated channels. This helps mitigate polarization-dependent intensity mismatch and s/p retardance\footnote{s/p retardance is the differential phase delay between the s- and p-polarized components of the beam.} in the upstream VLTI/NOTT optical path, which could otherwise affect the calibration of the interferometric outputs. The Wollaston is a custom \(30 \times 30~\mathrm{mm}\) MgF\(_2\) two-prism assembly designed for the \(3.0{-}4.0~\mu\mathrm{m}\) wavelength range, covering the NOTT \(L^{\prime}\)-band operating range.
    \smallbreak
    Following the Wollaston, the polarization-separated channels are spectrally dispersed by a filter-wheel-mounted CaF\(_2\)/ZnSe prism assembly, providing a low-resolution dispersive mode. The \(30 \times 30~\mathrm{mm}\) CaF\(_2\) and \(40 \times 40~\mathrm{mm}\) ZnSe prisms have nominal prism angles of \(40^\circ\) and \(16.5^\circ\), respectively. Additional dispersive elements could be accommodated in the filter wheel to enable future medium- and high-resolution modes, following the grism-based spectral-resolution concepts described by Dandumont et al. \cite{Dandumont2022} Rotation of the filter wheel is provided by an AML D42.2 ultra-high-vacuum stepper motor. This two-phase hybrid motor is specified for operation below \(1\times10^{-10}~\mathrm{mbar}\), making it suitable for in-vacuum mechanism control. A separate \(90~\mathrm{K}\) copper thermal strap, with length \(132~\mathrm{mm}\) and cross-sectional area \(90~\mathrm{mm^2}\), connects the motor to the base plate to provide a conductive thermal path.
    \smallbreak
    The spectrally dispersed, polarization-separated channels then enter the imager assembly, which comprises four ZnSe lenses that image the spectra onto the H2RG detector. The lenses have diameters of \(46.0~\mathrm{mm}\), \(42.0~\mathrm{mm}\), \(38.0~\mathrm{mm}\), and \(30.0~\mathrm{mm}\), and follow a mounting approach similar to that of the collimator assembly, using centering rings and spring preload to maintain lens positioning during cooldown. Since the detector is required to operate below \(50~\mathrm{K}\), its cold strap was dimensioned using a Simulink thermal model developed at the Australian National University (ANU). To maintain the detector at its operating temperature, the resulting design uses a copper cold strap of length \(161~\mathrm{mm}\) and cross-sectional area \(88~\mathrm{mm^2}\), together with a closed-loop heater power of \(2.72~\mathrm{W}\), which is within the nominal power-handling capability of the selected MP820 heaters.
    \smallbreak
    
    \begin{figure}[!ht]
    \centering
    \includegraphics[width=0.88\textwidth]{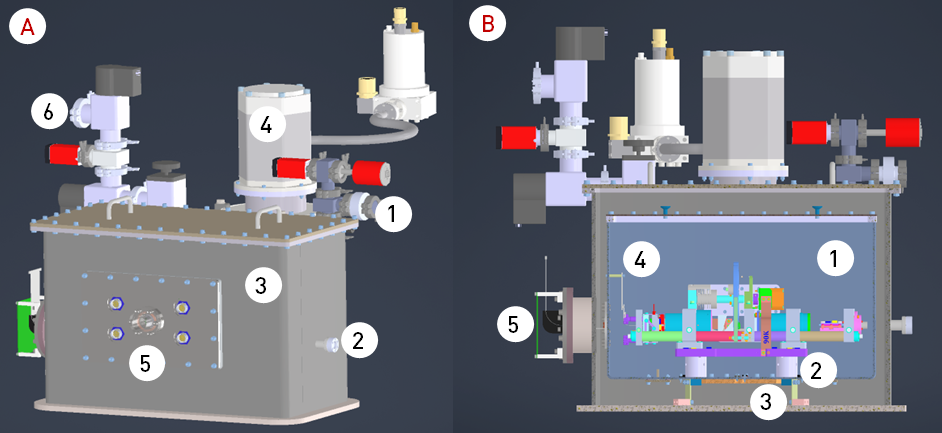}
    \caption{NOTT cryostat architecture. (A) External view showing the pressure sensors (1), entrance window (2), stainless-steel vacuum vessel (3), Cryomech PT805 pulse-tube cryocooler (4), electrical feedthrough (5), and interface to the turbomolecular pump station (6). (B) Section view showing the aluminium radiation shield (1), aluminium spacer (2), G10 blade (3), SIDECAR ASIC cold electronics (4), and SIDECAR feedthrough with MACIE warm electronics (5).}
    \label{cryostat}
    \end{figure}
    
    Built around a two-stage Cryomech PT805 pulse-tube cryocooler, the cryostat architecture provides the cooling power and thermal environment required for both the cold optics and the H2RG detector. The cryostat architecture in Fig.~\ref{cryostat} shows the stainless-steel vacuum vessel, aluminium radiation shield, base plate, and cold nulling camera. The H2RG is linked to the second stage of the cryocooler through its dedicated cold strap, while the aluminium radiation shield is thermally coupled to the cryocooler through copper thermal straps, with the main coupling to the first stage and an additional smaller coupling to the second stage. The cold nulling camera is mounted on the base plate via aluminium spacers attached to the shield. The shield is mechanically supported from the vessel by G10 blades, which are aligned with the natural deformation path to allow unconstrained deformation of the cold-optics interface. The design also includes four MP850 heaters and two PT100 temperature sensors on the shield to control the temperature distribution.

\section{SUMMARY AND FUTURE WORK}

    The NOTT cryostat is currently undergoing assembly, integration, and verification to validate the thermo-mechanical performance of the cold nulling camera, including cooldown stability, thermal interfaces, vacuum performance, and the operating temperatures of the cold optics and detector. The cryostat follows the ESO concept for small cryostats, with a volume of approximately \(47~\mathrm{L}\) and a total mass close to \(150~\mathrm{kg}\). Its thermal design uses a single aluminium radiation shield without MLI, electro-polished surfaces with emissivity \(<0.1\), and a two-stage Cryomech PT805 pulse-tube cooler. Steady-state thermal simulations predict cooling powers of \(23.3~\mathrm{W}\) at \(60~\mathrm{K}\) for the first stage and \(6.7~\mathrm{W}\) at \(40~\mathrm{K}\) for the second stage. 
    \smallbreak
    Initial cryogenic tests have demonstrated temperatures of approximately \(100~\mathrm{K}\) for the cold optics and \(30~\mathrm{K}\) for the detector, together with a vacuum pressure of \(8 \times 10^{-8}~\mathrm{mbar}\). The filter-wheel mechanism has also been validated at cryogenic temperature, supporting the selection of spectral-dispersion modes. Ongoing work focuses on completing the AIV campaign, improving the cold-optics operating temperature toward the \(90~\mathrm{K}\) requirement, verifying thermal stability during extended operation, and integrating the cold camera with the upstream NOTT optical system.

\label{sec:stat}

\acknowledgments 
SCIFY has received funding from the European Research Council (ERC); Award no. CoG - 866070 under the European Union's Horizon 2020 research and innovation program. This work received funding from the Hungarian NKFIH NKKP project No. K-147380. The authors S.E., T.A.S., and J.P.S. acknowledge support by the National Aeronautics and Space Administration (NASA) through grant 80NSSC23K1473.

\bibliography{report} 
\bibliographystyle{spiebib} 

\end{document}